\documentclass[traditabstract]{aa}
\usepackage{graphicx,epsfig,natbib,sidecap}
\usepackage{txfonts}
\usepackage{setspace}
\usepackage{threeparttable}

\usepackage{natbib}
\bibpunct{(}{)}{;}{a}{}{,} 

\begin{document}

\title{Doppler shift of hot coronal lines in a moss area of an active region}

\author{Neda Dadashi\inst{1,2}
  \and Luca Teriaca\inst{1}
  \and Durgesh Tripathi\inst{3}
  \and Sami K. Solanki\inst{1,4}
  \and Thomas Wiegelmann\inst{1}}

\offprints{N. Dadashi, \email{dadashi@mps.mpg.de}}

\institute{Max-Planck-Institut f\"{u}r Sonnensystemforschung, 37191 Katlenburg-Lindau, Germany
\and Institut f\"{u}r Geophysik und extraterrestrische Physik, Technische
Universit\"{a}t Braunschweig, Mendelssohnstr. 3, D-38106 Braunschweig, Germany
\and Inter-University Center for Astronomy and Astrophysics, Post Bag 4, Ganeshkhind, Pune 411007,
India \and School of Space Research, Kyung Hee University, Yongin, Gyeonggi-Do,
446-701, Korea}

\date{Received 3 September 2012 / Accepted  7 November 2012 }

\abstract
{ The moss is the area at the footpoint of the hot (3 to 5 MK)
loops forming the core of the active region where emission is believed to
result from the heat flux conducted down to the transition region from
the hot loops. Studying the variation of Doppler shift as a function
of line formation temperatures over the moss area can give clues on the
heating mechanism in the hot loops in the core of the active regions.
We investigate the absolute Doppler shift of lines formed at temperatures
between 1 MK and 2 MK in a moss area within active region NOAA 11243
using a novel technique that allows determining the absolute Doppler shift of
EUV lines by combining observations from the SUMER and EIS spectrometers.
The inner (brighter and denser) part of the moss area shows roughly
constant blue shift (upward motions) of 5 km s$^{-1}$ in the temperature
range of 1 MK to 1.6 MK. For hotter lines the blue shift decreases and
reaches 1 km s$^{-1}$ for Fe~{\sc xv} 284 \AA\ ($\sim$ 2 MK). The measurements
are discussed in relation to models of the heating of hot loops.
The results for the hot coronal lines seem to support the quasi-steady
heating models for non-symmetric hot loops in the core of active regions.}


\authorrunning{Dadashi, Teriaca, Tripathi, Solanki, Wiegelmann}

\titlerunning{Doppler shift of hot coronal lines in a moss region}

\maketitle

\section{Introduction}
Active regions dominate the solar emission at EUV and X-ray wavelengths
whenever they are present on the solar surface. Moreover they are the
sources of most of the solar energetic phenomena. As such, active regions
are a privileged target for studies of the solar activity of magnetic
origin. Thus, studying the motions and flows over
active regions have the potential of setting observational constraints on
models of coronal heating and provide some clues to solve this problem
\citep{Doschek-etal-2007, Hara-etal-2008, DelZanna2008, Brooks-warren2009, warren-etal-core-AR-2010}.

Recent Observations from Hinode \citep{Culhane2007} show two different types
of active region loops: warm loops \citep[$\sim$1 MK,][]{Ugarte-Urra2009}
and hot loops \citep[$>$ 2 MK,][]{Brooks-etal-2008}. There is evidence that the
warm loops are multi-stranded structures impulsively heated by storms
of nanoflares \citep{Warren-etal-2003, klimchuk2006, Tripathi-etal-2009, Ugarte-Urra2009, klimchuk2009}.

However, in the case of hot coronal loops, there is observational support for
both steady heating \citep{Antiochos-etal-2003, warren-winebarger-etal-2008, Winebarger-etal-2008, Brooks-warren2009, warren-etal-core-AR-2010, Winebarger-etal-2011}
and impulsive heating \citep{tripathi-mason-2010, Bradshaw-klimchuk-2011, Tripathi-klim-mas-2011, viall-klimchuk2012}.
 Because of the unresolved nature of hot coronal loops
\citep{Tripathi-etal-2009} it is not possible to isolate a single loop
and study its characteristics. One of the alternatives is to study the
footpoints of the hot coronal loops, in the so-called "moss" areas.
The moss is a bright reticulated feature observed in the EUV and was first
described by \citet{berger-depon-flet99}. They concluded that the moss is
emission due to heating of low-lying plasma by thermal conduction
from overlying hot loops.
Thus, studying the spatial and temporal characteristics of observables such
as brightness \citep{Antiochos-etal-2003},
Doppler shifts \citep{Doschek-etal-2008, DelZanna2008, Brooks-warren2009, Tripathi-mason-klim-2012, Winebarger_2012_sub},
line widths \citep{Doschek-etal-2008, Brooks-warren2009},
electron densities \citep{Fletcher_bart1999, Tripathi-etal-2008, tripathi-young-2010, Winebarger-etal-2011}
and temperature structure \citep{tripathi-young-2010} in the moss could
provide an important constraint on the heating mechanism operating in the hot
core loops.

Measuring Doppler shift in the moss region is a powerful diagnostic tool to
distinguish between steady heating and impulsive heating models
\citep{Brooks-warren2009, DelZanna2008, Tripathi-mason-klim-2012}.
Since the detected Doppler shifts in the above mentioned studies were in
general not significantly larger than the associated uncertainties
the exact behaviour of the flows at temperatures above 1 MK is still
debated and needs to be studied with higher accuracy.

\citet{Antiochos-etal-2003} found that the intensity inside the moss
region varies very weakly (only 10\%) over periods of hours. They also found
that the magnetic field inside the moss regions changes very slowly.
This was the first indication that the heating in the moss region is not due
to discrete impulsive flare-like events,
but to either steady heating or to low energy, high-frequency events
that approximate a quasi-steady process.
Heating in symmetric loops foresees no motion and therefore no
Doppler shifts because the loops are supposed to be in static equilibrium.
Any kind of asymmetry (like pressure difference between the two photospheric
footpoints of the loops or in the heating and/or cross section of the loops),
can generate steady flows \citep{Noci1981, Boris-mariska-82, Mariska-boris83}.
Flows produced in this manner by \citet{Boris-mariska-82} have small
velocities of few hundred m s$^{-1}$ unless the asymmetries are extreme
\citep{orlando-etal-1995b, Winebarger-etal-2002, Patsourakos-etal-2004}.
The most common flows produced by these asymmetries are uni-directional
\citep[like siphon flow,][]{Noci1981, Craig1986}.
This means one of the loop legs should show blue shift
and the other red shift. In the case of pressure difference between
footpoints, overpressure at one footpoint of the loop drives an upflow moving
along the entire loop and drains at the opposite footpoint. The flow
accelerates with height due to density decrease in accordance with the
continuity equation\footnote{$\frac{1}{A} \frac{\partial}{\partial s}(nvA)=0$,
where n, v, A, and s are density, velocity, cross section, and length
coordinate along the 1D loop, respectively.} \citep{Aschwanden2004book}.

Heating at both footpoints of the loops can also produce flows if it is
significantly concentrated toward the footpoints
\citep[and references therein]{klimchuk-karp-antio2010}. The source of the
heating could be either truly steady or the frequency of the impulsive
heating could be sufficiently high that a steady approximation would be
valid. The flows are driven because of thermal non-equilibrium.
Since the loop has localized heating at both sides, evaporative upflows
(blue shifts) should occur from both ends of the loop.

Impulsive heating models consider the active region loops to be composed of
many hundreds of small elemental unresolved strands that are randomly heated
by storms of nanoflares \citep{Cargill1994}. Some of the strands could show
blue shifts (upflows) due to chromospheric evaporation and some of the strands
could show red shifts (downflows) due to cooling and condensation of the
evaporated plasma. Simulations done by \citet{Patsourakos-klim2006} have
revealed that upflows are faster, fainter and have a shorter lifetime than
downflows. The computed line profiles from these strands are found to have
a red shifted core with an enhanced blue wing. The red shifts are predicted
to decrease with increasing temperature with sufficiently hot lines being
blue shifted (Bradshaw \& Klimchuk, priv. communication).

\citet{Brooks-warren2009} measured the Doppler shift, non-thermal width
and temporal variation of the Fe~{\sc xii} 195 \AA\ line over a moss
region. They obtained a small red shift of about 2 to 3 km s$^{-1}$ with
almost no change in Doppler shift and non-thermal velocity with time. They
concluded that their result verifies quasi-steady heating models. The value
they used as a reference for measuring the Doppler shift, was obtained by
averaging over the whole raster. However, this is not an accurate way to
measure small velocities because if the emission outside of the moss has a
non-zero absolute Doppler shift, which is rather likely, their reported
result for the moss Doppler shift could change considerably.

\citet{Tripathi-mason-klim-2012}, using a more reliable wavelength
calibration method developed by \citet{young-etal-2012}, studied the
Doppler shifts in the temperature range of 0.7 - 1.6 MK.
They obtained blue shifts below 2 km s$^{-1}$ for both their coolest
(Fe~{\sc ix} 197 \AA) and hottest (Fe~{\sc xiii} 202 \AA) lines, with an
estimated accuracy of 4 to 5 km s$^{-1}$. They conclude that their result is
in agreement with predictions of both steady and impulsive heating scenarios.
Since the uncertainties in their work are quite larger than their measured
velocities, to reveal the real direction of the motions in the moss area,
a more accurate Doppler shift measurements with smaller uncertainties is
needed.

In the present paper, we
have concentrated on studying bulk flows by measuring Doppler shifts in the
moss region using a high precision method which is based on simultaneous
observations from the Extreme ultraviolet Imaging Spectrometer
\citep[EIS,][]{Culhane2007} aboard Hinode and the Solar Ultraviolet
Measurement of Emitted Radiation \citep[SUMER,][]{Wilhelm95} spectrometer
aboard the Solar and Heliospheric Observatory (SoHO). The implication of
these results on the heating of hot loops rooted in moss areas is
discussed.

\section{Data analysis}
\begin{figure}[t]
  \centering
  \resizebox{10.cm}{!}{\includegraphics{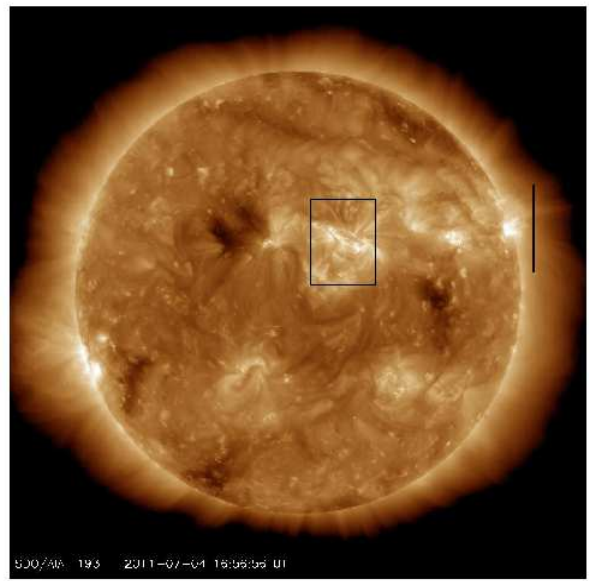}}
  \caption{Fe~{\sc xii} 193~\AA\ AIA image. The area observed on the disk
    by EIS is outlined by the black box. The black perpendicular line on the right
    side of the image represents the EIS slit position off-limb. Only the
    bottom quarter of the off-limb slit is used for data analysis. The area
    scanned by SUMER is shown in a closer view in Figure \ref{chap8_f1}.}
 \label{chap8_f00}
 \end{figure}

Coordinated SUMER and EIS data (HOP193) were taken on 4th July 2011, between
15:50 to 18:53 UTC over NOAA active region 11243 near disk center.
A full disk image by the Atmospheric Imaging Assembly
\citep[AIA,][]{Lemen2012-aia} aboard the Solar Dynamics Observatory (SDO)
around the time of our observation is shown in Figure \ref{chap8_f00}.

\subsection{SUMER data: Doppler shift of Mg~{\sc x}}
Since 1995 the normal incidence spectrograph, SUMER, operates between
450~\AA\ to 1610~\AA wavelength range. This powerful UV instrument is
designed to investigate the bulk motions of plasma in the chromosphere,
transition region, and low corona. The spatial resolution of
SUMER across and along the slit is 1 arcsec and 2 arcsec, respectively.
The spectral scale
\citep[$\approx$ 43 m\AA/pix at 1240 \AA\ in the first order of diffraction,][]{Wilhelm97-1, Lemaire97} is accurate
enough\footnote{in case of high signal to noise ratios} to
measure the Doppler shift of lines down to 1 km s$^{-1}$. Corrections for
wavelength-reversion, dead-time,  flat-field, and detector
electronicdistortion are applied to SUMER raw spectral images.

We use SUMER to measure the absolute Doppler shift of one coronal
line (Mg~{\sc x} 625 \AA) observed in the second order of diffraction
around 1250 \AA. SUMER does not have an on-board calibration source, hence we
obtain the wavelength scale by using a set of chromospheric lines
from neutrals and singly ionized atoms (for instance, C~{\sc i} lines with
formation temperatures of 10000 K.).
These lines are known to have negligible or small average Doppler shifts
 \citep{Hassler91, Samain:1991}. The wavelength calibration has three main steps:
 identifying the lines by using a preliminarily wavelength scale, fitting
 the Gaussian curves to the spectral lines to find the exact pixel position
 of the lines, performing a polynomial fit to obtain the dispersion relation.

The SUMER data that we have used here consist of a
sit-and-stare\footnote{The slit stays fixed in space and let the Sun rotates beneath.}
sequence near disk center over a moss area within active region NOAA 11243.
The acquired spectra (1245 \AA\ to 1255 \AA\ containing the Mg~{\sc x} 625 \AA\
line, log T/[K]=6.00) are obtained by exposing for 75 s through the
1 $\times$ 300 arcsec$^2$ slit. The sequence consists of 48 spectra and
results in a small, 10 arcsec wide, drift scan by solar rotation.
 \begin{figure}
  \centering
  \resizebox{9.cm}{!}{\includegraphics{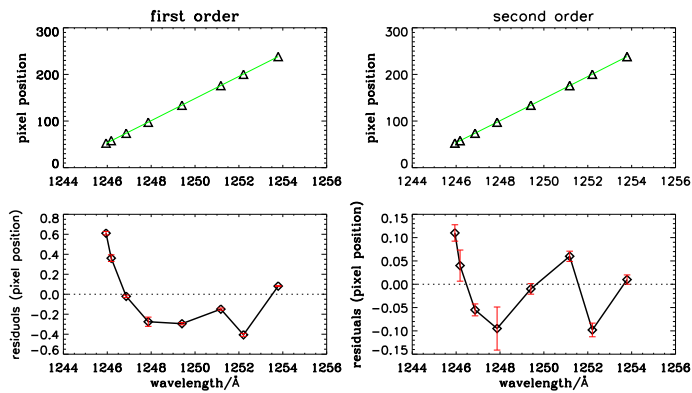}}
  \caption{Left column: top panel shows the first order polynomial fit to
           pixel positions versus rest wavelengths of 8 suitable reference
           lines around the Mg~{\sc x} 625 \AA\ line. Lower panel represents
           the residuals of the fitting. The residuals are large, up to 0.6
           pixels ($\sim$ 6 km s$^{-1}$).
           Right column: top and lower panels represent a second order polynomial fit
           to pixel positions versus rest wavelengths and the residuals of the
           fitting, respectively. The residuals lie within $\pm$ 0.1 pixels
           (about 1 km s$^{-1}$).}
 \label{chap8_new_fig}
 \end{figure}
We have checked the data to be sure that there are no significant
instrumental drifts in the position of the spectral lines (either along the
slit or as a function of time across the raster). A high signal to noise
spectrum is obtained by averaging over the region of interest. Then, by
choosing eight suitable reference lines (from relatively strong and unblended
lines of neutral or singly ionized atoms) we have performed the wavelength
calibration. A linear fit of their positions in pixels versus
their laboratory wavelengths results in residuals of less than $\pm$ 0.6
pixels (see left column of Figure \ref{chap8_new_fig}). The residuals clearly
show the need for a second-order polynomial fit that leaves residuals of $\pm$
0.1 pixels (less than 2 km s$^{-1}$). This is demonstrated in right column of
Figure \ref{chap8_new_fig}. \citet{dadashi2011} also used a second order
polynomial fit to obtain the velocity of the same line over a quiet Sun area.
We believe the need for the higher order fit comes from residual errors in
the correction of the electronic distortion of the detector image.

To obtain the Doppler shift map of the Mg~{\sc x} 625 \AA\ line, we have
used 624.967 \AA\ as rest wavelength. This is the average between the
values of (624.965 $\pm$ 0.003) \AA\ given by \citet{Dam99a} and
(624.968 $\pm$ 0.007) \AA\ given by \citet{PetJud99}.
The reasons for this selection of the above rest wavelength are described in
detail in \citet{dadashi2011}.

\subsection{EIS data: analysis and co-alignment}
The EIS instrument produces high-resolution stigmatic spectra in
the wavelength ranges of 170 to 210~\AA\ and 250 to 290~\AA.
The instrument has 1 arcsec spatial  pixels and 0.0223~\AA\ spectral
pixels. More details are given in \citet{Culhane2007} and
\citet{Korendyke2006}.

\begin{figure*}
  \centering
  \resizebox{13.cm}{!}{\includegraphics[angle=90]{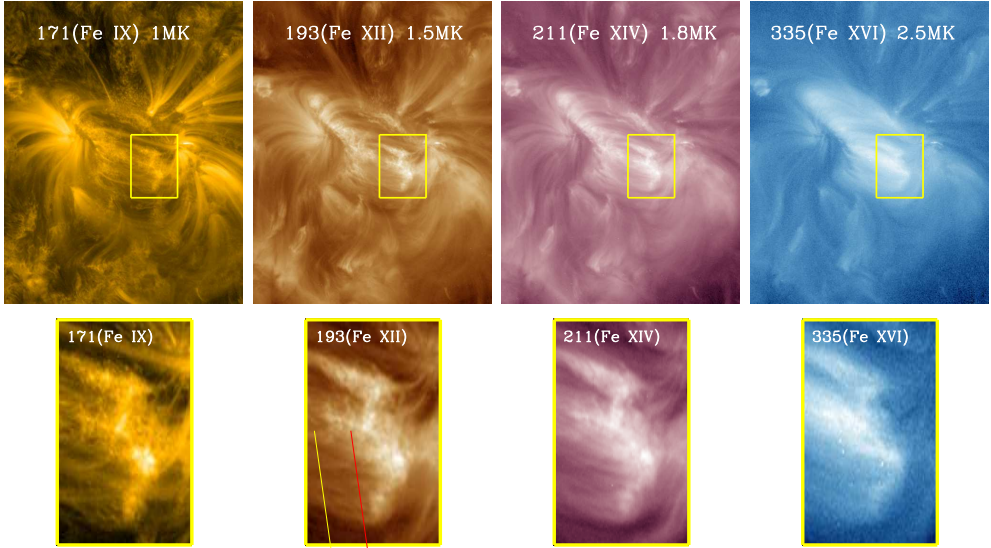}}
  \caption{ The NOAA active region 11243 is shown in different wavelength bands
            of AIA/SDO (top panels). The images are taken on 4th July 2011
            at 17:00 UTC. The yellow boxes show the moss area studied
            in this work. Lower panels are blow ups of the corresponding
            yellow boxes.}
 \label{chap8_f0}
 \end{figure*}

EIS data consists of one raster scan (1 arcsec step size and 30 s exposure
time) of the active region taken nearly simultaneously to the SUMER data,
followed by one sit-and-stare sequence taken above the limb.
Both observations were
obtained using the 1 arcsec wide slit. The area of study (NOAA active
region 11243) is shown in different wavelength bands of AIA/SDO in
Figure \ref{chap8_f0}. The over-plotted yellow box on each panel of this
Figure shows the moss area that we had studied here.
The eis\_prep.pro routine, which is the standard EIS data reduction
routine available in the SolarSoft (SSW) package, is used.
This routine subtracts the dark current, removes the cosmic rays
and hot pixels, and does radiometric calibration. The slit tilt and orbital
variation effects are removed from the data by using the SSW
eis\_wave\_corr.pro routine. The orbital variation caused by the thermal
effects on the instrument does not play an important role in our study,
since the technique we employ is based on the measurement of the line
separations (the distance between the line of which we want to measure the
shift and a reference line), a quantity that does not change during the
relatively short duration of our observations
\citep[for further details see][]{dadashi2011}.

\begin{figure}
  \centering
  \resizebox{9.cm}{!}{\includegraphics[angle=90]{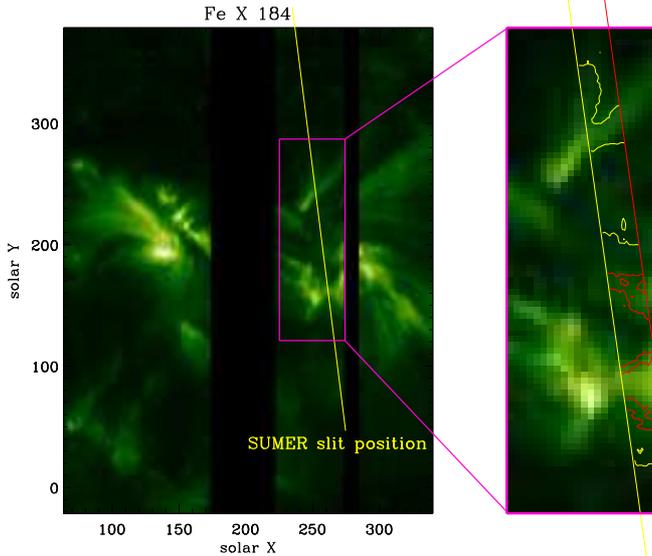}}
  \caption{Left panel: the position of the SUMER slit at the start
           of the drift scan over-plotted on the EIS Fe~{\sc x} 184 \AA\
           intensity image raster. The SUMER slit crosses the moss area that we
           have studied in this paper. Right panel: blow up of the pink
           box in the left panel. Intensity contours of
           Mg~{\sc x} 625 \AA\ are plotted on top of the intensity map of the
           Fe~{\sc x} 184 \AA\ line.}
 \label{chap8_f1}
 \end{figure}

After removing the offset between the SW and LW detectors, we co-align the
two instruments (SUMER and EIS) by using a pair of radiance maps obtained
in lines formed at similar temperatures: Mg~{\sc x} 625 \AA\
(SUMER) and Fe~{\sc x} 184 \AA\ (EIS). The left panel of Figure
\ref{chap8_f1} shows the position of the SUMER slit at the beginning of the
drift scan over-plotted on the EIS Fe~{\sc x} 184 \AA\ image raster.
The SUMER slit crosses the moss area that we have studied in this
paper. The right panel of Figure \ref{chap8_f1} is a blow-up of the pink box
shown in the left panel. The area scanned by SUMER is placed between red
and yellow inclined lines. Intensity contours of Mg~{\sc x} 625 \AA\
plotted on top of the intensity map of Fe~{\sc x} 184 \AA\ allow us to
co-align the image with an accuracy of roughly 1 arcsec.

\subsection{Method of measuring absolute Doppler shifts}\label{methode}
Since there are no suitable cool chromospheric lines in the EIS wavelength
range, it is not possible to use the same calibration technique as with SUMER.

One of the alternatives is to measure "relative" Doppler shifts of hot
coronal lines by comparing to observations of quiet Sun areas that are
present in the same raster scan containing the target active region.
To obtain "absolute" Doppler shifts with this technique,
\citet{young-etal-2012} have used the average absolute velocities
in the quiet Sun measured by \citet{PetJud99} using SUMER spectra.
However, there is some degree of uncertainty and arbitrariness in
identifying what is quiet Sun near an active region, leading to a
substantial uncertainty of the absolute velocity.

\citet{dadashi2011} have introduced a novel technique to obtain "absolute"
Doppler shifts of hot coronal lines using simultaneous observation of EIS and
SUMER. The technique is based on two important assumptions:
\begin{itemize}
  \item First, above the limb of quiet regions without obvious structures,
        the spectral lines have, on average, no Doppler shift because the
        motions out of the plane of sky cancel out on average.
  \item Second, Mg~{\sc x} 625 \AA\ (SUMER) and
        Fe~{\sc x} 184 \AA\ (EIS) lines, which have similar
        formation temperature, have the same average Doppler shift in the
        common area of study\footnote{Reasons are described in \citet{dadashi2011}.}.
\end{itemize}
Following \citet{dadashi2011}, we consider the EIS Fe~{\sc x} 184 \AA\
line as the reference line and obtain all other coronal line velocities
respect to this line.

\begin{equation}
\overline {\delta v}  = \frac{c}{{\lambda _0 }}\left( {\overline {\Delta \lambda }  - \overline {\Delta \lambda _{0ff} } \,\left( {1 + \frac{{\overline {v^{Ref} } }}{c}} \right)} \right).
\label{equ:1}
\end{equation}

\begin{figure}
  \centering
  \resizebox{10.cm}{!}{\includegraphics{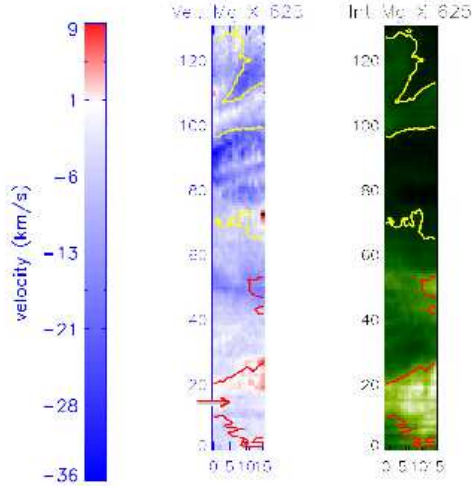}}
  \vspace{-1.5cm} \caption{Intensity and velocity maps of Mg~{\sc x} 625 \AA\
           obtained by the sit-and-stared SUMER observations. Contours of
           Fe~{\sc x} 184 \AA\ are plotted on both maps. The area that is
           marked by the red arrow corresponds to region \textbf{b} in the moss area
           (defined in the left panel of Figure \ref{chap8_}). The average
           Doppler velocity of Mg~{\sc x} 625 \AA\ is of $-$6.6 km s$^{-1}$
           inside region \textbf{b} and of ($-$5.50 $\pm$ 0.55) km s$^{-1}$
           over the whole area shown here.}
 \label{chap8_f2}
 \end{figure}

Then, the absolute Doppler shifts are obtained as

\begin{equation}
\overline {v} = \overline {\delta v} + \overline {v^{Ref}},
\label{equ:2}
\end{equation}
where $\overline {\delta v}$ and $\overline {v}$ are the relative
(to Fe~{\sc x}) and absolute average Doppler shift of each line and
$\overline {v^{Ref}}$ is the average Doppler shift of the Mg~{\sc x} line
measured by SUMER (see Fig. \ref{chap8_f2}), assumed to be equal to that
of the reference line
(Fe~{\sc x}). $c$ and $\lambda _0$ are the speed of light in vacuum and
the rest wavelength of the line of which we want to measure the shift.
Since $\lambda _0$ is not used to calculate the shift of the spectral
line due to the Doppler effect, it does not need to be known with high accuracy.

$\overline {\Delta \lambda }$ and $ \overline {\Delta \lambda _{0ff} }$ are
the average of the distribution of wavelength separation of the two
lines (the line of which we want to measure the shift and the reference line)
on disk and off-limb, respectively. In this work we use this technique to
measure the absolute Doppler shift of hot coronal lines in the moss region
of NOAA active region 11243. To have a higher signal-to-noise ratio, before
fitting the line profiles, we have binned the on-disk spectra over two raster
positions and over two pixels along the slit (2$\times$2 binning). In the
case of off-limb spectra we have used a binning of 3$\times$20 (3 in solar X
and 20 in solar Y direction).

This way, maps of relative Doppler shift were first obtained
for Fe~{\sc xi} 188 \AA, Fe~{\sc xii} 192 \AA,
Fe~{\sc xiii} 202 \AA, Fe~{\sc xiv} 270 \AA, and
Fe~{\sc xv} 284 \AA\ lines. The absolute Doppler shift map of
Fe~{\sc x} 184 \AA\ was obtained by imposing as rest wavelength a
value that gives the same velocity pattern as Mg~{\sc x} 625 \AA\ in
the common observed area. After this, all relative Doppler maps were easily
converted into absolute Doppler maps.

\subsection{Moss identification}
\begin{figure}
  \centering
  \resizebox{9.cm}{!}{\includegraphics{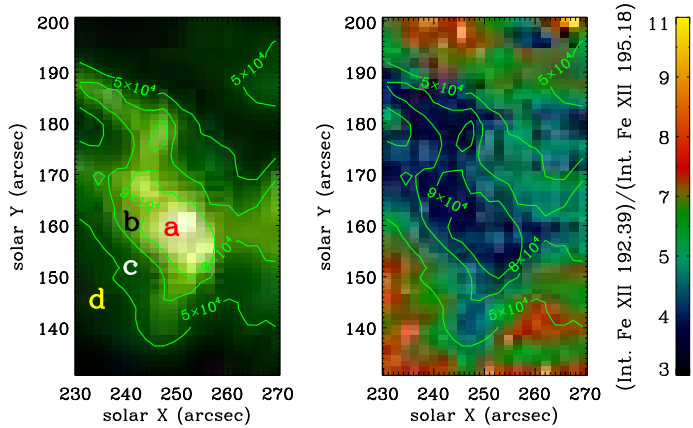}}
  \caption{Left panel: Intensity map of Fe~{\sc xii} 192 \AA.
           Right panel: The intensity ratio of Fe~{\sc xii} 192 \AA\
           and Fe~{\sc xii} 195.2 \AA\ lines. Smaller ratios correspond
           to higher electron densities. The identified moss area (regions
           \textbf{a} and \textbf{b}) lie in the regions with higher
           density, as expected. Contours of intensity of the
           Fe~{\sc xii} 192 \AA\ line are plotted on both maps.}
 \label{chap8_}
 \end{figure}

Figure \ref{chap8_} shows three different contours of the intensity of
the Fe~{\sc xii} 192 \AA\ line dividing the whole area of study into
4 different regions. We consider the middle intensity contour as a
threshold to identify moss (regions \textbf{a} and \textbf{b} together,
as marked in the left panel of Fig. \ref{chap8_}).

\begin{figure}
  \centering
  \resizebox{9.cm}{!}{\includegraphics{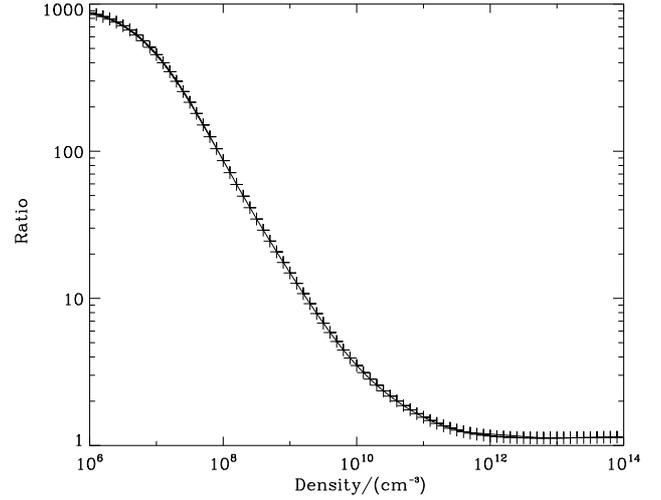}}
  \caption{Intensity ratio of Fe~{\sc xii} 192 \AA\ and
           Fe~{\sc xii} 195.18 \AA\ lines as a function of density.
           Smaller intensity ratios correspond to higher densities.}
 \label{chap8__}
 \end{figure}

The intensity ratio of Fe~{\sc xii} 192.39 \AA\ and Fe~{\sc xii} 195.18 \AA\
is density-sensitive. Using the CHIANTI atomic database \citep{CHIANTI2012},
we have derived the intensity ratio of these two lines as a function of
density (Fig. \ref{chap8__}). Smaller intensity ratios correspond to higher
densities. Right panel of Fig. \ref{chap8_} shows the map of the intensity
ratio of the above Fe~{\sc xii} lines.
Brighter intensity contours of Fe~{\sc xii} 192.39 \AA\ line, which
determine the moss region, are plotted on both maps. The identified moss
region is characterized by higher density, as expected. The average
electron density for regions \textbf{a}, \textbf{b}, \textbf{c}, and
\textbf{d} are listed in Table \ref{table:chap8_1}.
The average moss density (average of regions \textbf{a} and \textbf{b})
is about 6.6 $\times$ 10$^9$ cm$^{-3}$, which is a bit larger than
4.2 to 5.3 $\times$ 10$^9$ cm$^{-3}$ reported by \citet{Fletcher_bart1999}.
On the other hand, \citet{tripathi-young-2010} obtained larger electron
densities of about 10 to 30 $\times$ 10$^9$ cm$^{-3}$ using the
Fe~{\sc xii} 195.12 \AA\ and 186.88 line ratio.
However, we should consider that the densities obtained by Fe~{\sc xii} lines are
overestimated in high density regions. The new Fe~{\sc xii} calculations by
\citet{DelZanna12-chianti-fexii} show that there was some problem with
the atomic data which has now been resolved.
\begin{table}
\caption{The Fe~{\sc xii} density average over the different regions
defined in the left panel of Figure \ref{chap8_}.}
\label{table:chap8_1}
\centering
 \begin{tabular*}{5.5cm}{lc}  
  \hline  \hline  & \mbox{}\\[-1.5ex]
  region & average density (cm$^{-3}$)        \\
  \hline & \mbox{}\\[-1.5ex]
  a      &  7.27 $\times$ 10$^9$ \\
  b      &  6.31 $\times$ 10$^9$ \\
  c      &  4.84 $\times$ 10$^9$ \\
  d      &  3.39 $\times$ 10$^9$ \\
  \hline & \mbox{}\\[-1.5ex]
 \end{tabular*}
 \end{table}

\section{Result and discussion}
Using the technique introduced by \citet{dadashi2011}, and summarized in
Section \ref{methode}, we have obtained the line of sight Doppler shifts
of hot coronal lines in a moss area of the active region NOAA 11243.
In the first step of our study, we have defined the moss area by intensity
contours of Fe~{\sc xii} 192 \AA. Based on these contours, four different
areas are defined in that region (left panel of Fig. \ref{chap8_}).
Table \ref{table:chap8_2} lists the Doppler shift values for each of these
regions with the corresponding uncertainties. Positive and negative values
of velocity represent downflows (red shifts) and upflows (blue shifts),
respectively.
\begin{table*}
\caption{The average Doppler shift of hot coronal lines in the different regions
defined in the left panel of Fig. \ref{chap8_}. Positive and negative values
of velocity represent downflows (red shifts) and upflows (blue shifts),
respectively.}
\label{table:chap8_2}
\centering
 \begin{tabular*}{10.5cm}{c c c c c c}  
\hline \hline & \mbox{}\\[-1.5ex]
 line (log(T/[K]))  & \multicolumn{4}{c}{~~~~~~$\overline {v}$} &$~~~~~~\delta(\overline {v})$\\
                    & \multicolumn{4}{c}{~~~~~~[km~s$^{-1}$]}   & ~~~~~~[km~s$^{-1}$]  \\
                    &~~~~~~  \#~a & \#~b & \#~c &  \#~d & \\
  \hline & \mbox{}\\[-1.5ex]
~Fe~{\sc xi} 188.216 (6.04)  & $-$2.83 & $-$5.14 & $-$5.51 & $-$6.45 & 2.27 \\
                                                       \\
Fe~{\sc xii} 192.394 (6.15) & $-$3.70 & $-$5.06 & $-$5.50 & $-$8.57 & 2.51 \\
~~~~S~{\sc x} 264.233 (6.15)& $-$4.48 & $-$5.39 & $-$5.05 & $-$7.31 & 2.06 \\
                                                       \\
Fe~{\sc xiii} 202.044 (6.20)& $-$3.78 & $-$4.67 & $-$3.93 & $-$8.57 & 1.96 \\
                                                       \\
Fe~{\sc xiv} 270.519 (6.26) & $-$1.04 & $-$1.82 & $-$1.60 & $-$6.81 & 2.15 \\
                                                       \\
Fe~{\sc xv} 284.160 (6.32)  & $-$1.52 & $-$0.36 & $-$0.46 & $-$7.75 & 2.08 \\
  \hline & \mbox{}\\[-1.5ex]  \\
 \end{tabular*}
 \end{table*}

 \begin{figure}
  \centering
  \resizebox{9.cm}{!}{\includegraphics[angle=90]{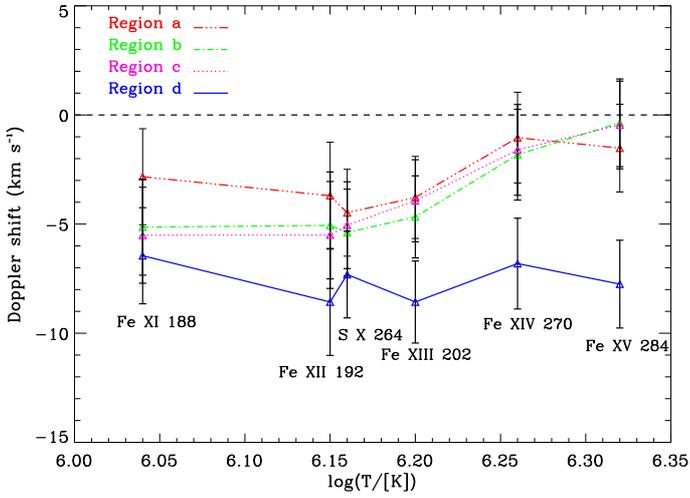}}
  \caption{The average Doppler shift of hot coronal lines in different
           regions defined in the left panel of Fig. \ref{chap8__}, plotted
           as a function of formation temperature. Fe~{\sc xii} and S~{\sc x}
           have nearly the same formation temperature, but are plotted a bit
           separated for better readability. Positive and negative values
           of velocity represent downflows (red shifts) and upflows
           (blue shifts), respectively.}
 \label{chap8_f4}
 \end{figure}

The error reported in this Table $\delta(\overline {v})$ is obtained by
error propagation analysis of Eqs. \ref{equ:1} and \ref{equ:2}.
The error on $\overline {\Delta \lambda_{0ff}}$ is given by the width
$\sigma$ of the off-limb line-separation distribution. Since the on-disk line
separation distribution is also broadened by the different Doppler speeds
characterizing the two lines at different temperatures, the error for
$\overline {\Delta \lambda}$ is assumed to be equal to that off-limb.
This leads to the same error for a given line in all different regions on disk.
 \begin{figure}
  \centering
  \resizebox{4.cm}{!}{\includegraphics[angle=90]{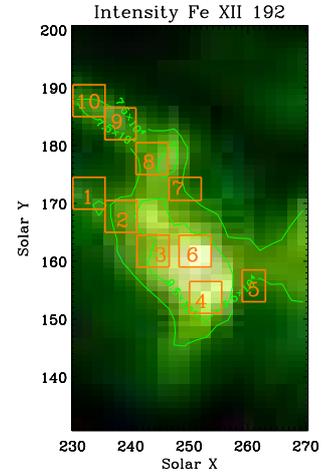}}
  \caption{ Ten different small square areas selected to study the Doppler
            velocities therein, overplotted on a map of
            Fe~{\sc xii} 192 \AA\ intensity. Green contours are intensity
            contours of Fe~{\sc xii} 192 \AA. }
 \label{chap8_f5}
 \end{figure}
 \begin{figure}
  \centering
  \resizebox{9.cm}{!}{\includegraphics[angle=90]{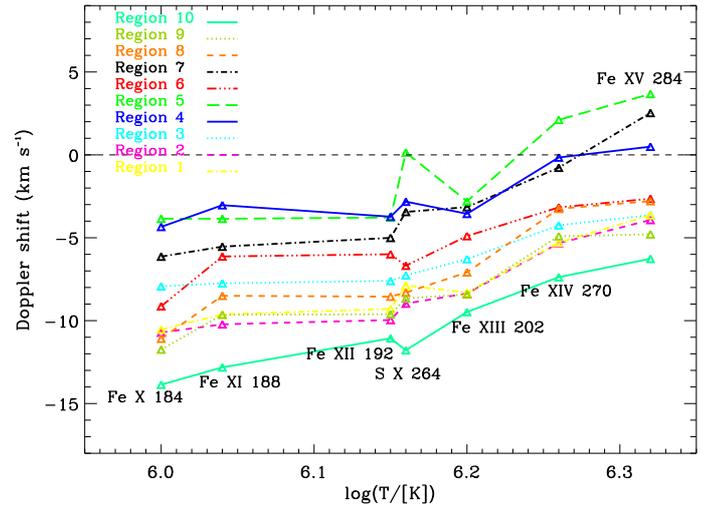}}
  \caption{ The Doppler shift as a function of temperature for the
            ten boxes defined in Fig. \ref{chap8_f5}. Fe~{\sc xii}
            and S~{\sc x} have the same formation temperature but are plotted
            a bit separated for better readability.}
 \label{chap8_f6}
 \end{figure}

Figure \ref{chap8_f4} shows the average Doppler shift for the four regions
defined by intensity contours of the Fe~{\sc xii} 192 \AA\ as a function
of temperature. The Doppler shifts of the lines in the temperature range from
1 to 1.6 MK over regions \textbf{a}, \textbf{b}, and \textbf{c} show constant
blue shifts or upward motions of about 5 km s$^{-1}$. For hotter lines
the upward motions decrease down to about 1 km/s for Fe~{\sc xv}. Region
\textbf{a}, which corresponds to the highest electron densities and
intensities of the Fe~{\sc xii} 192 \AA\ line,
shows a somewhat smaller blue shift than
the other regions for temperatures below 1.8 MK. Region \textbf{d}, that
has the smallest intensity and density values, shows almost constant and
stronger blue shift of about 8 km s$^{-1}$, in the temperature range from
1 to 2 MK.
 \begin{figure*}
  \centering
  \resizebox{18.cm}{!}{\includegraphics[angle=90]{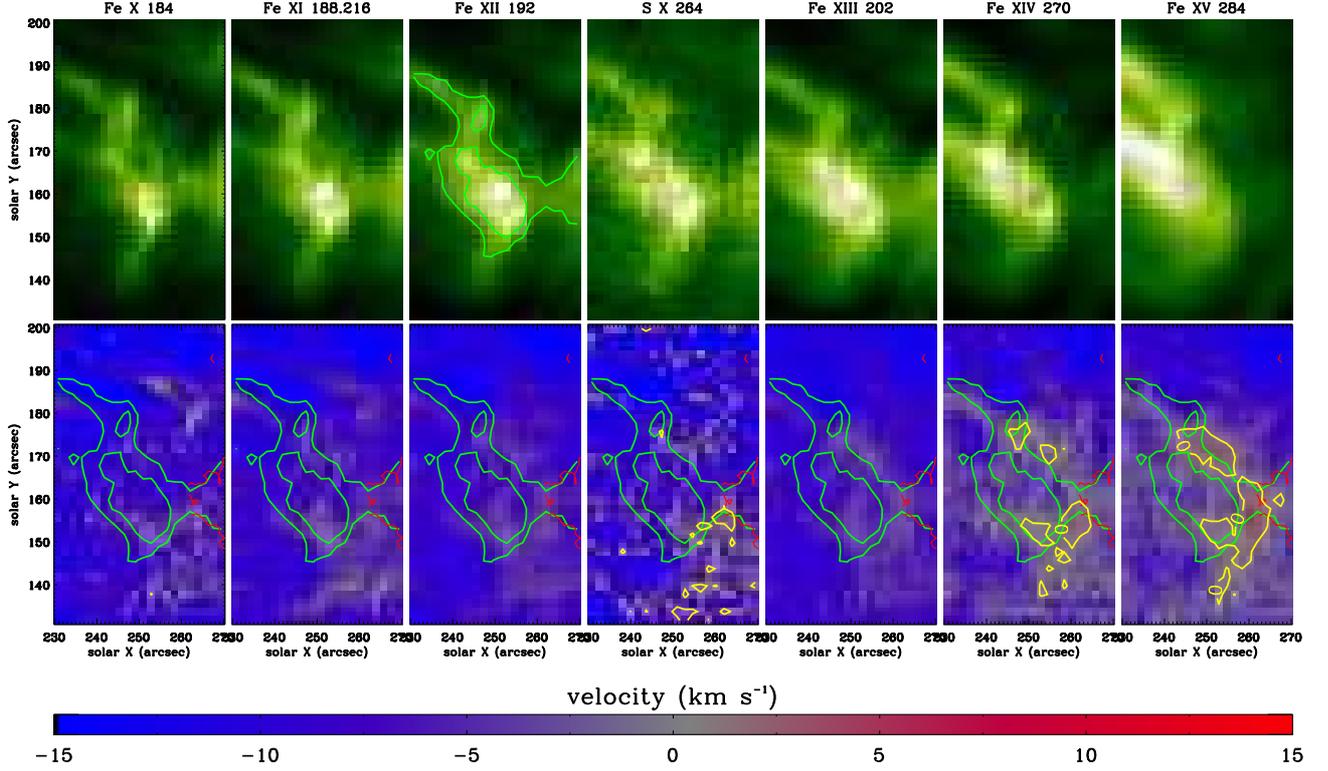}}
  \caption{Intensity (upper panels) and velocity maps (lower panels)
           for Different ions with different formation temperatures.
           Green contours are intensity contours of Fe~{\sc xii} 192 \AA\
           plotted on top of intensity (top row) and velocity maps (bottom row)
           of different ions. Yellow contours on velocity maps outline
           red shifted pixels. Red contour is the brightest area which is
           common in both EIS and SUMER field of views and is used for velocity calibration. }
 \label{chap8_f3}
 \end{figure*}

For hotter lines (1.6 to 2 MK) there is almost no prevalent motion in
regions \textbf{a}, \textbf{b}, and \textbf{c} (0 to 2 km s$^{-1}$, with
velocities never diverging by more than 2$\sigma$ from zero). Region
\textbf{d} at the same temperature shows strong blue shift.

To have a better understanding of the behaviour, we have selected ten small
boxes at different locations of the area of interest (shown in Fig.
\ref{chap8_f5}) and calculated the average absolute Doppler shifts
inside each box. Figure \ref{chap8_f6} displays the average absolute Doppler
shifts of different ions as a function of their formation temperatures in
the ten selected areas.

\begin{figure}
  \centering
  \vspace{-2cm} \resizebox{9.cm}{!}{\includegraphics{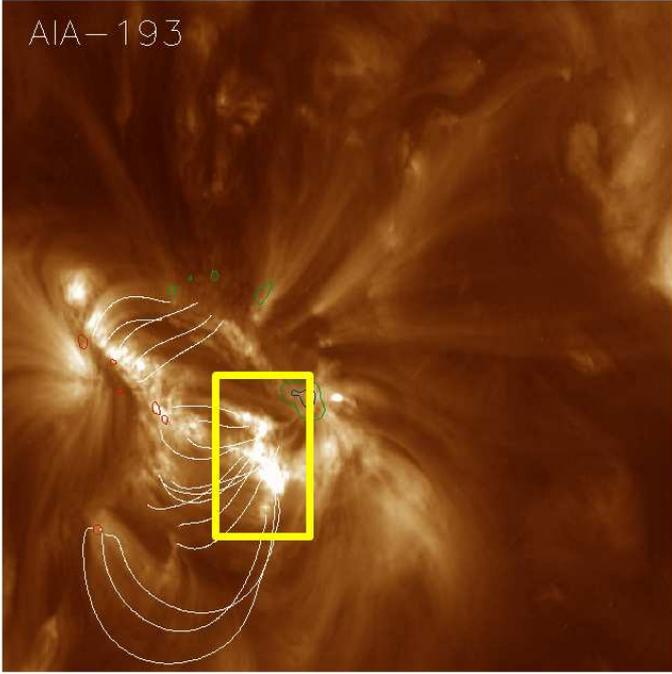}}
  \vspace{0.5cm} \caption{Magnetic field lines extrapolated from photospheric
                          field and plotted over Fe~{\sc xii} 193 \AA\
                          image of AIA/SDO. Red contours show positive and blue
                          and green contours show negative polarities. The
                          yellow frame represents the region of the moss
                          study in this work. }
 \label{chap8_magnetic}
 \end{figure}

The same general behaviour of velocity is observed with respect to
temperature in all regions. Doppler shifts are essentially constant
up to temperatures of about 1.6 MK. Above this value, upward velocities
decrease and almost stop at a temperature of about 2 MK except for the loop
legs and the parts of the moss with lower electron densities
(e.g. regions 1, 2, 3, 9, and 10).
The inner (brighter and denser) part of the moss shows smaller upflows
(regions 4 and 5). In contrast, regions 1, 2 and 8, 9, which are on the
legs of the hot coronal loops (as suggested also by magnetic field
extrapolations, see Fig. \ref{chap8_magnetic}), show larger blue shift
(upward motions) of about 10 km s$^{-1}$ between 1 MK and 1.6 MK.
This suggest that as the plasma rises higher along the loop leg, it
accelerates. This conclusion is further illustrated by region 10, which lies
further up in the loop leg and is associated with higher upflows of about 13
to 14 km s$^{-1}$). In all regions, colder lines sense stronger upflows
(the colder, the more shifted to the blue).

Hotter lines like Fe~{\sc xiv} 270 \AA\ and Fe~{\sc xv} 284 \AA\ in the
brighter and denser part of moss reveal no significant motions (the only
exception is region 5 that shows a small red shift (downward motion) of
about 3 km s$^{-1}$ at this temperature).
The intensity and corresponding Doppler shift maps of all these hot coronal
lines are plotted in Fig. \ref{chap8_f3}.
The red shifted area of Fe~{\sc xiv} and Fe~{\sc xv} lies within the yellow
contours overplotted on the corresponding velocity maps.

Our results for Fe~{\sc xii} are based on the unblended Fe~{\sc xii} line at
192 \AA. However, we have also obtained the velocity from the
Fe~{\sc xii} 195.120 line which is
one of the brightest lines recorded by EIS. This line is blended by
Fe~{\sc xii} 195.180 \AA\ in its red wing. The density sensitive ratio,
195.180/195.120, increases with increasing density. Considering both
components, performing a double Gaussian fit will result in
redder velocities than the other unblended Fe~{\sc xii}
lines\footnote{http://msslxr.mssl.ucl.ac.uk:8080/eiswiki/Wiki.jsp?page=EisDiscussion}
\citep{young-fexii2009}. Here we find that considering a double Gaussian
profile for Fe~{\sc xii} 195 \AA\ on average results in a velocity
redder than Fe~{\sc xii} 192 \AA\ by about 2.2 km s$^{-1}$ .

To have a more precise idea about the orientation of the hot coronal loops,
we have extrapolated the photospheric field into the solar corona by using
photospheric vector magnetograms obtained by the Helioseismic and Magnetic
Imager \citep[HMI,][]{HMI-SDO2012} aboard Solar Dynamics Observatory (SDO).
The nonlinear force-free coronal magnetic field extrapolation technique used
here is described in detail by \citet{Wiegelmann2012}. Figure
\ref{chap8_magnetic} shows the magnetic field lines plotted over an AIA/SDO
image taken at 17:00 UTC, which lies within the time frame of our study. The
field of view for applying nonlinear coronal magnetic field extrapolation
(the whole window of Fig. \ref{chap8_magnetic}) is chosen in order to
fulfill consistency criteria:
\begin{itemize}
  \item The vector magnetogram should be almost perfectly flux balanced.
  \item The field of view should be large enough to cover weak field surrounding
        of the target active region.
\end{itemize}

The field lines shown in Fig. \ref{chap8_magnetic} better reveal the
prevailing direction of the hot loops than can be seen in the 211 and
335 AIA channels. Note that the loops emerging from the moss extend away
from it in two directions. These are the short loops extending North-East
from the upper part of the moss. These correspond to the loops well visible
in Fe~{\sc xiv} and Fe~{\sc xv}, whose legs are sampled by boxes 1, 2, 8, 9,
and 10 in Fig. \ref{chap8_f5}. The results of the magnetic field
extrapolation show another set of loops which is directed the South-East
starting from the moss (more visible in cooler lines like  Fe~{\sc ix} 171 \AA).
Thus, the Fe~{\sc xiv} and Fe~{\sc xv}
red shifted regions visible in Fig. \ref{chap8_f3} (outlined by
the yellow contours) appear located in the areas free of line-of-sight
contamination from plasma higher up along the loop legs. Since the emission
from the loop legs is clearly blue shifted, we may speculate that the near
zero blue shifts observed over the moss may actually be a combination of
red shift at the footpoint and blue shift along the loops. Both red and
blue shifts are too small to leave a signature in the line profile
sufficiently strong to disentangle the two velocity components in a single
line profile at the EIS spectral resolution. However, the possibility
that the red shifted region seen in Fe~{\sc xv} is due to an independent
phenomenon, cannot be entirely excluded.

 \begin{figure*}
  \centering
  \resizebox{14.cm}{!}{\includegraphics{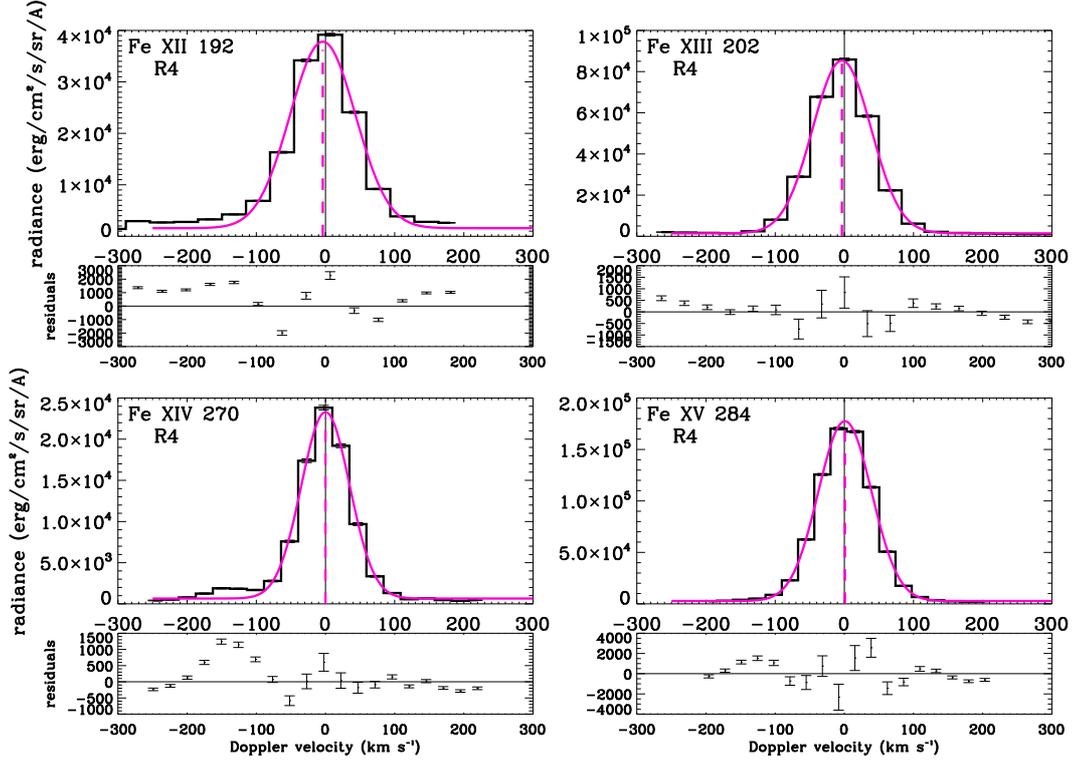}}
  \caption{Spectral profiles of Fe~{\sc xii}, Fe~{\sc xiii}, Fe~{\sc xiv},
           and Fe~{\sc xv} obtained over region 4 (defined in Fig.
           \ref{chap8_f5}).}
 \label{chap8_f7}
 \end{figure*}
Impulsive heating models that consider a group of unresolved strands heated
by periodic events like nanoflares, predict
red shift (downward motion) on both legs of the loops in $\sim$1 MK lines
and more complex profiles with an extended blue wing in very hot lines
such as Fe~{\sc xvii} \citep{klimchuk2006, Patsourakos-klim2006}.

\begin{figure*}
  \centering
  \resizebox{14.cm}{!}{\includegraphics{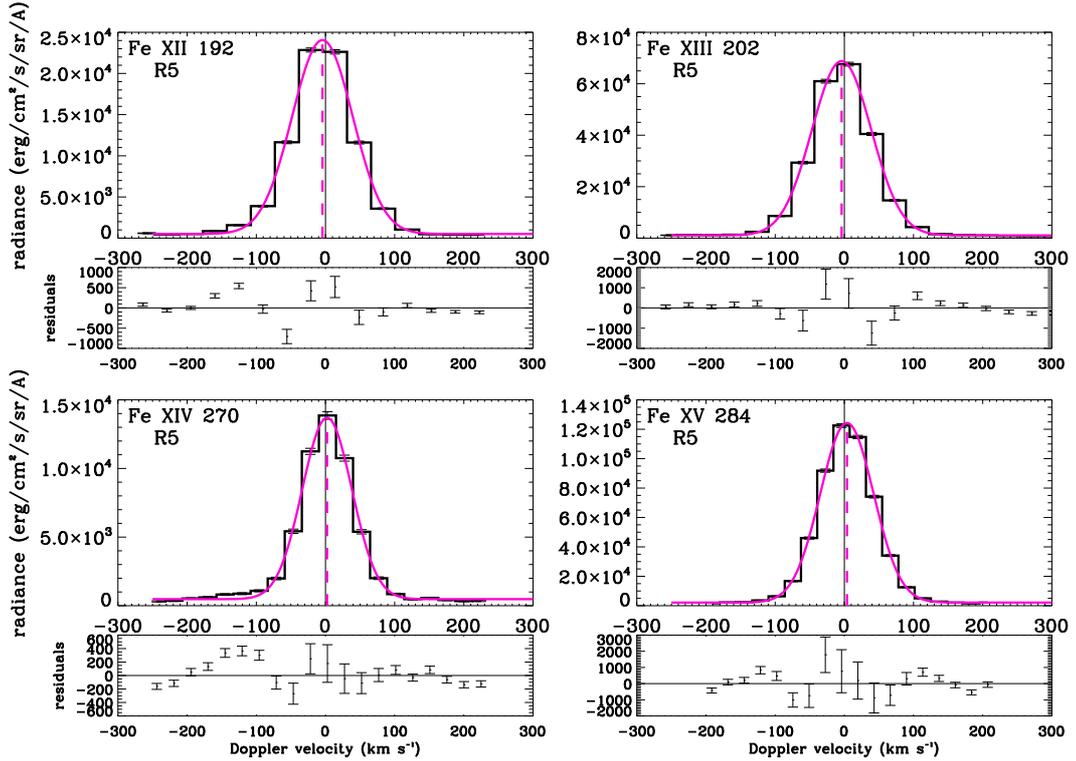}}
  \caption{Spectral profiles of Fe~{\sc xii}, Fe~{\sc xiii}, Fe~{\sc xiv},
           and Fe~{\sc xv} obtained over region 5 (defined in Fig.
           \ref{chap8_f5}).}
 \label{chap8_f8}
 \end{figure*}

\begin{figure*}
  \centering
  \resizebox{14.cm}{!}{\includegraphics{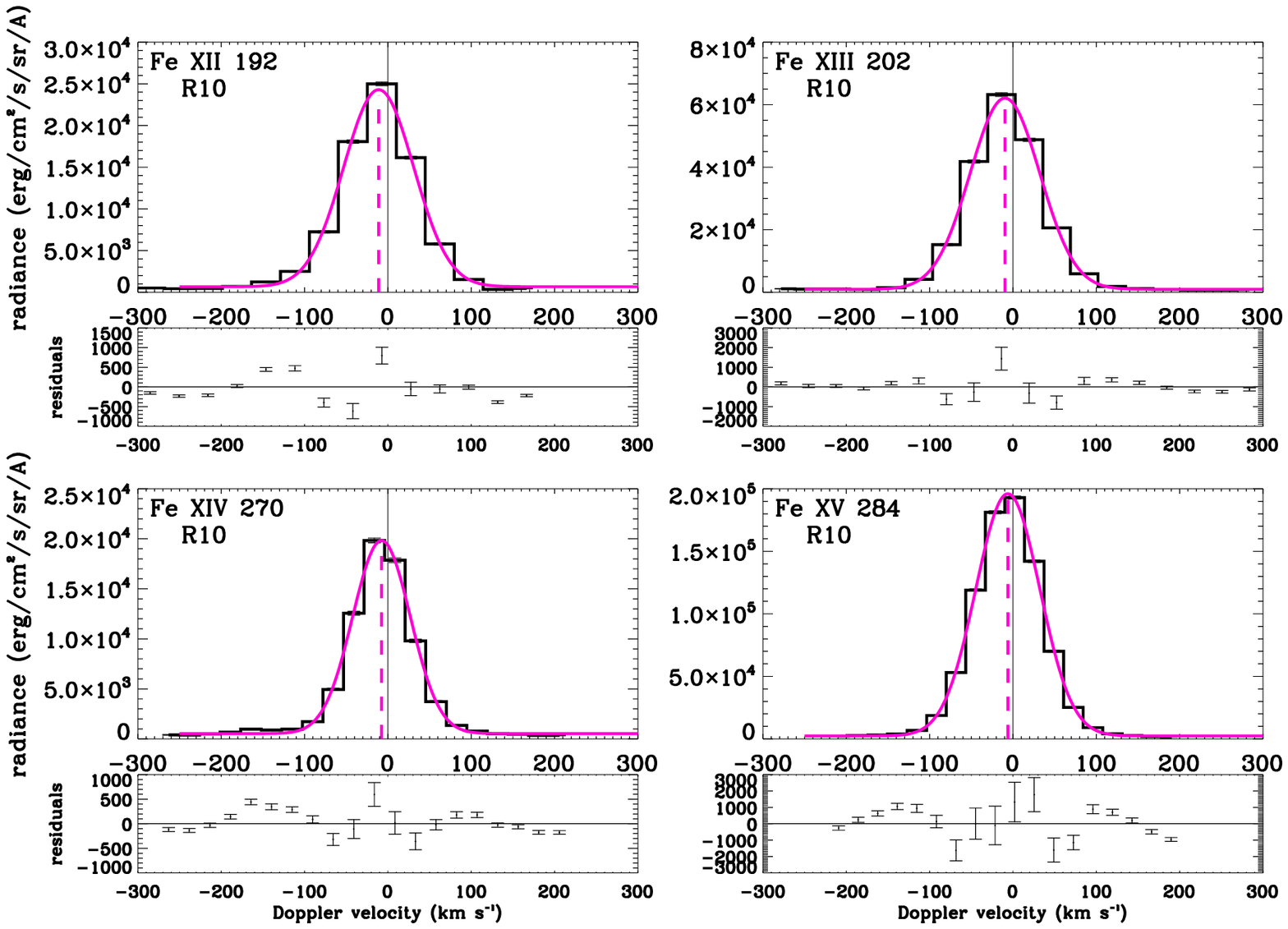}}
  \caption{Spectral profiles of Fe~{\sc xii}, Fe~{\sc xiii}, Fe~{\sc xiv},
           and Fe~{\sc xv} obtained over region 10 (defined in Fig.
           \ref{chap8_f5}).}
 \label{chap8_f10}
 \end{figure*}

Our Doppler shift measurements inside the moss region do not show the
predicted red shift of $\sim$1 MK lines and the increasing blue shift of
lines at increasingly higher temperatures, but rather the opposite. We
clearly see blue shifted emission in lines formed between 1 and 1.6 MK and
almost zero velocity at the Fe~{\sc xv} formation temperature.
Moreover, we have studied the line profiles associated with some of these
ten regions to look for possible asymmetries. Figures \ref{chap8_f7}
to \ref{chap8_f10} show the spectral line profiles for
Fe~{\sc xii} 192 \AA, Fe~{\sc xiii} 202 \AA,
Fe~{\sc xiv} 270 \AA, and Fe~{\sc xv} 284 \AA\ over the selected
regions 4, 5, and 10, respectively. A single Gaussian fit is performed on
all line profiles. Residuals of each fit are also plotted
right under the panel with the line profile.

The spectral profiles seem to be well represented by a Gaussian and no
specific asymmetry is detected (residuals are less than 2 to 3\% of the
amplitudes of the line profiles).
Fe~{\sc xiv} 270 \AA\ and Fe~{\sc xv} 284 \AA\ line profiles seem
to be weakly blended with Mg~{\sc vi} 270.394 \AA\ and
Al~{\sc ix} 284.015 \AA\ in their blue wings (the small bumps in the
residual plots). After taking out the effect of these blends, all line
profiles are highly symmetric \citep[see also][]{Tripathi-lineprofile-12}.
If profiles with extended blue wings are actually present at the footpoints
of hot loops, they are only observable in lines much hotter that Fe~{\sc xv}.

Measurements of density and thermal properties of moss observed by EIS have
shown no significant change over several hours \citep{tripathi-young-2010}.
This leads the authors to suggest that the heating of the hot coronal loops
associated to the moss might be quasi-steady. \citet{Brooks-warren2009} have
shown that the amount and variability of shifts and widths of coronal lines
is small and appears consistent with steady heating models.
However, we note that variability of density, flows and line widths were
studied on a spatially averaged region and any small scale variations may get
averaged out. Steady heating models of perfectly symmetric coronal loops
predict no bulk motions since the conductive flux from the corona is
balanced by radiative cooling \citep{Mariska-boris83, klimchuk-karp-antio2010}.
Impulsive heating, on the other hand, predicts definite plasma flows, which
are temperature dependent. The relationship between velocity and temperature
after spatial averaging
may not be straight forward, as different strands may be evolving completely
independently and providing mixed observational signatures.

The near absence of Doppler shifts in the hotter lines (Fe~{\sc xiv} and
Fe~{\sc xv}) and, in general, the absence of red shifted emission in the
observed moss region, along with the observed symmetric line profiles,
seems to be consistent with quasi-steady heating of non-symmetric loops.
Steady heating driven by pressure difference between the footpoints
of the loops predicts the flows to be accelerated with height.
For non-symmetric loops, this effect will produce siphon-like flows in which
one footpoint will be red shifted and the other would be blue shifted.
Unfortunately, we can not check and compare the flows at the other footpoint
of these loops. In fact, due to the fact that the Hinode spacecraft suffered
from seasonal eclipses at that time of the year, the other footpoint is
missing in the EIS image raster (Figure \ref{chap8_f1}).

In support of quasi-steady heating of hot loops, we have found the Doppler
shift to increase as we go higher in the loop leg (Figures \ref{chap8_f5}
and \ref{chap8_f6}). However, observations of a larger sample of moss regions
in lines covering higher temperatures than Fe~{\sc xv} would be necessary
to consolidate our findings.

\section{Summary}
\begin{itemize}
  \item By combining SUMER and EIS data, with a new technique developed by
        \citet{dadashi2011}, we have measured the absolute Doppler shift
        of hot coronal lines (Fe~{\sc x} 184 \AA, Fe~{\sc xi} 188 \AA,
        Fe~{\sc xii} 192 \AA, Fe~{\sc xiii} 202 \AA, Fe~{\sc xiv} 270 \AA,
        and Fe~{\sc xv} 284 \AA) in the moss of an active region.
  \item The moss is identified as the region where Fe~{\sc xii} 192 \AA\
        intensity corresponds to electron densities above about
        6.6 $\times$ 10$^9$ cm$^{-3}$.
  \item The inner (brighter and denser) part of the moss area shows roughly a
        constant blue shift (upward motions) of 5 km s$^{-1}$ in the
        temperature range of 1 MK to 1.6 MK. For hotter lines the blue shift
        decreases, down to 1 km s$^{-1}$ for Fe~{\sc xv} 284 \AA.
        Absolute Doppler shift maps are obtained using the \citet{dadashi2011}
        technique. The general dependence of the velocity on temperature
        seems to be the same everywhere in the map. In all regions, the
        colder the line the stronger the blue shift.
  \item In general, the inner (brighter and denser) part of the moss shows
        smaller blue shifts, whereas the legs of hot coronal loops with
        footpoints in the moss, show larger blue shift.
  \item Our velocity measurements inside the moss region do not show the
        red shifts predicted by the impulsive loop heating model
        \citep{klimchuk2006, Patsourakos-klim2006}. Also, we have found the
        line profiles to be highly symmetric, which is in contrast with the
        predictions of impulsive heating models.
  \item The near absence of motions seen in the hotter lines and, in
        general, the absence of red shifted emission in the observed moss
        region, as well as the observed symmetric line profiles, and higher
        velocities in higher parts of the loop legs seem to be consistent
        with quasi-steady heating models for non-symmetric loops.
\end{itemize}

\begin{acknowledgements}

The SUMER project is financially supported by DLR, CNES, NASA, and the ESA
PRODEX programme (Swiss contribution). SoHO is a mission of international
cooperation between ESA and NASA.
Hinode is a Japanese mission developed and launched by ISAS/JAXA,
collaborating with NAOJ as a domestic partner, and NASA and STFC (UK) as
international partners. Scientific operation of the Hinode mission is
conducted by the Hinode science team organized at ISAS/JAXA. This team
mainly consists of scientists from institutes in the partner countries.
Support for the post-launch operation is provided by JAXA and NAOJ (Japan),
STFC (U.K.), NASA, ESA, and NSC (Norway).
Data are courtesy of NASA/SDO and the AIA and HMI science teams.
This work has been partially supported by WCU grant No. R31-10016 funded by
the Korean Ministry of Education, Science and Technology.
ND acknowledges a PhD fellowship of the International Max Planck
Research School on Physical Processes in the Solar System and Beyond.
DT would like to thank MPS for supporting his travel and providing
excellent hospitality.
\end{acknowledgements}

\bibliographystyle{aa}

\bibliography{dadashi}

\end{document}